\documentclass[aps,prb,amsmath,amssymb,twocolumn,a4paper,floatfix,showpacs]{revtex4-1}
\usepackage{graphicx}
\usepackage{float}
\usepackage{hyperref}
\usepackage{color}

\begin{document}
\title{Slope-Reversed Mott Transition in Multiorbital Systems}
\author{Aaram J. Kim}
\altaffiliation[Present Address:]{
Institut f\"ur Theoretische Physik,
Goethe-Universit\"at Frankfurt am Main,
Max-von-Laue-Stra\ss e 1,
60438 Frankfurt am Main,
Germany
}\affiliation{Department of Physics and Astronomy and Center for Theoretical Physics, Seoul National University, Seoul 151-747, Korea}

\author{MooYoung Choi}
\affiliation{Department of Physics and Astronomy and Center for Theoretical Physics, Seoul National University, Seoul 151-747, Korea}
\author{Gun Sang Jeon}
\email{gsjeon@ewha.ac.kr}
\affiliation{Department of Physics, Ewha Womans University, Seoul 120-750, Korea}

\begin{abstract}
We examine finite-temperature phase transitions
in the two-orbital Hubbard model with different bandwidths
by means of the dynamical mean-field theory combined with the continuous-time quantum Monte Carlo method.
It is found that there emerges a peculiar slope-reversed first-order Mott transition between the orbital-selective Mott phase and the Mott insulator phase in the presence of Ising-type Hund's coupling.
The origin of the slope-reversed phase transition is clarified by the analysis of the temperature dependence of the energy density.
It turns out that the increase of Hund's coupling lowers the critical temperature of the {slope-reversed} Mott transition. 
Beyond a certain critical value of Hund's coupling the first-order transition turns into a finite-temperature crossover.
We also reveal that the orbital-selective Mott phase exhibits frozen local moments in the wide orbital, which is demonstrated by the spin-spin correlation functions.
\end{abstract}


\maketitle

\section{INTRODUCTION}
\label{sec:intro}

Coexistence of strongly and weakly correlated electrons has been one of the
intriguing subjects in condensed matter physics.
Materials in which more than one orbital is active near the Fermi level have
exhibited interesting properties and their main origin is believed to be the
coexistence of electrons with different degrees of correlations~\cite{Georges2013}.
In the multiorbital system, correlations between electrons and Hund's coupling
have been known to show rich phenomena in the presence of the orbital degree of
freedom.
In case that the degeneracy between active orbitals is lifted by the difference
of their bandwidths~\cite{Arita2005,Ferrero2005,Koga2005,Koga2005a,Koga2004,Koga2002,Lee2011a,Greger2013,Knecht2005,Lee2010a,Liebsch2005}
or crystal-field splitting~\cite{Werner2007,Hafermann2012,Jakobi2013,deMedici2009,Kita2011},
the degree of effective correlations in each orbital becomes different.
One prominent consequence of different degrees of correlations is the
orbital-selective Mott phase (OSMP), where electrons in some orbitals are totally
localized due to the Mott physics while other orbitals are still occupied by
itinerant electrons~\cite{Nakatsuji2000,Anisimov2002}.
Here Hund's coupling tends to intensify the difference between orbitals~\cite{deMedici2011,Georges2013}.

The coexistence of strongly and weakly correlated
electrons is also believed to play an important role in two-dimensional materials including strong spatial fluctuations~\cite{Sakai2009,Zhang2007a}.
In such a system spatial correlations and corresponding momentum-space anisotropy of
correlations are the key elements to host the coexistence~\cite{Gull2010}.
Thanks to the recent numerical developments in the cluster dynamical mean-field theory (DMFT)~\cite{Hettler1998,Hettler2000,Lichtenstein2000,Kotliar2001,Maier2005,Kotliar2006,Tremblay2006}, it is known that spatial fluctuations
modify qualitatively finite-temperature behaviors of the correlation-driven
metal-insulator transitions; this has been revealed by the comparison with the
single-site DMFT~\cite{Georges1996} neglecting
spatial fluctuations.
Spatial correlations turn out to reduce greatly the ground-state entropy of the
paramagnetic Mott insulator (MI) at low temperatures and accordingly, the
itinerant bad metallic phase dominates in the region of relatively high temperatures near the transition~\cite{Park2008}.

It is natural to anticipate such prominent changes in finite-temperature
transitions for multiorbital systems.
In spite of extensive studies~\cite{Knecht2005,Lee2010a,Liebsch2005,Jakobi2013,Werner2007,Hafermann2012,Liebsch2004,Liebsch2003,Biermann2005,Liebsch2006},
the temperature dependence of the transitions in the two-orbital Hubbard model
still lacks a thorough understanding.
The principal purpose of this work is
to investigate the finite-temperature nature
of the transitions in two-orbital systems with emphasis on the effects of Hund's coupling.

In this paper we investigate the two-orbital Hubbard model by the DMFT
combined with the continuous-time quantum Monte Carlo (CTQMC) method.
In the model we find the slope-reversed Mott transition in the presence of
Ising-type Hund's coupling for two orbitals of different bandwidths.
We also observe that the drastic changes in the phase transition between the
OSMP and the MI phase are induced by the variation of the Hund's coupling strength.
The analysis of the hysteresis behavior of local magnetic moments determines the
location of the critical end points, which reveals that the critical temperature
tends to reduce as the Hund's coupling is increased. 
Eventually the hysteretic behavior disappears at a certain value of Hund's
coupling and the system exhibits only a crossover between the OSMP and the MI phase.
We also compute the spin-spin correlation function for both orbitals and find the formation of the local frozen moments for itinerant wide-orbital electrons in the OSMP.

This paper is organized as follows: In Sec. II we give a brief description of
the two-orbital Hubbard model and the numerical method. 
Section~III is devoted to the presentation of the numerical results, which
include finite-temperature phase diagrams, spectral functions, hysteresis of
local magnetic moments, energy densities, effects of Hund's coupling, and spin-spin correlation functions.
The results are summarized in Sec.~IV.

\section{MODEL AND METHODS}
\label{sec:modelmethod}

We consider the Hamiltonian 
\begin{eqnarray}
	\mathcal{H} &=& -\sum_{\langle ij\rangle \alpha\sigma} t_\alpha(\hat{c}^{\dagger}_{i\alpha\sigma}\hat{c}^{}_{j\alpha\sigma} + h.c.)  - \mu\sum_{i\alpha\sigma}\hat{n}^{}_{i\alpha\sigma}\nonumber\\
	&&+
	U\sum_{i\alpha}\hat{n}^{}_{i\alpha\uparrow}\hat{n}^{}_{i\alpha\downarrow}+\sum_{i\sigma\sigma'}(U'-J\delta_{\sigma\sigma'})\hat{n}^{}_{i1\sigma}\hat{n}^{}_{i2\sigma'},
	\label{eqn:hamiltonian}
\end{eqnarray}
for two orbitals $\alpha=1$ and $2$.
Here, $\hat{c}^{}_{i\alpha\sigma}$($\hat{c}^{\dagger}_{i\alpha\sigma}$) is the
annihilation(creation) operator of an electron with spin $\sigma$ at site $i$
and orbital $\alpha$.
In each orbital
electrons move on the infinite-dimensional Bethe lattice
corresponding to a noninteracting semi-circular density of states (DOS),
$\rho^0_\alpha(\omega) = (2/\pi D_\alpha)\sqrt{1-(\omega/D_\alpha)^2}$,
with the half bandwidth $D_\alpha=2 t_\alpha$
and interact with each other via
the intra- and inter-orbital Coulomb interactions $U$ and $U'$
and Hund's coupling $J$.
We investigate the half-filled system with chemical potential $\mu=3U/2 -5J/2$,
and also choose $D_2 = 2D_1$ and $U'=U-2J$.
The half bandwidth $D_1$ of the narrow orbital is taken as an energy unit throughout this paper.
Here we disregard spin-flip and pair-hopping terms, which is appropriate for the study
of the anisotropic Hund's coupling model.
Our model serves as a natural generalization of the Ising-spin Kondo lattice model 
which is useful in interpreting experimental results for various materials, e.g., pyrochlore oxides~\cite{Udagawa2012,Chern2013} and URu$_2$Si$_2$~\cite{Sikkema1996}.
Such physics is understood mainly in terms of the large-anisotropy effects on the localized moments.

\begin{figure*}
	\centering
	\includegraphics[width=0.86\textwidth]{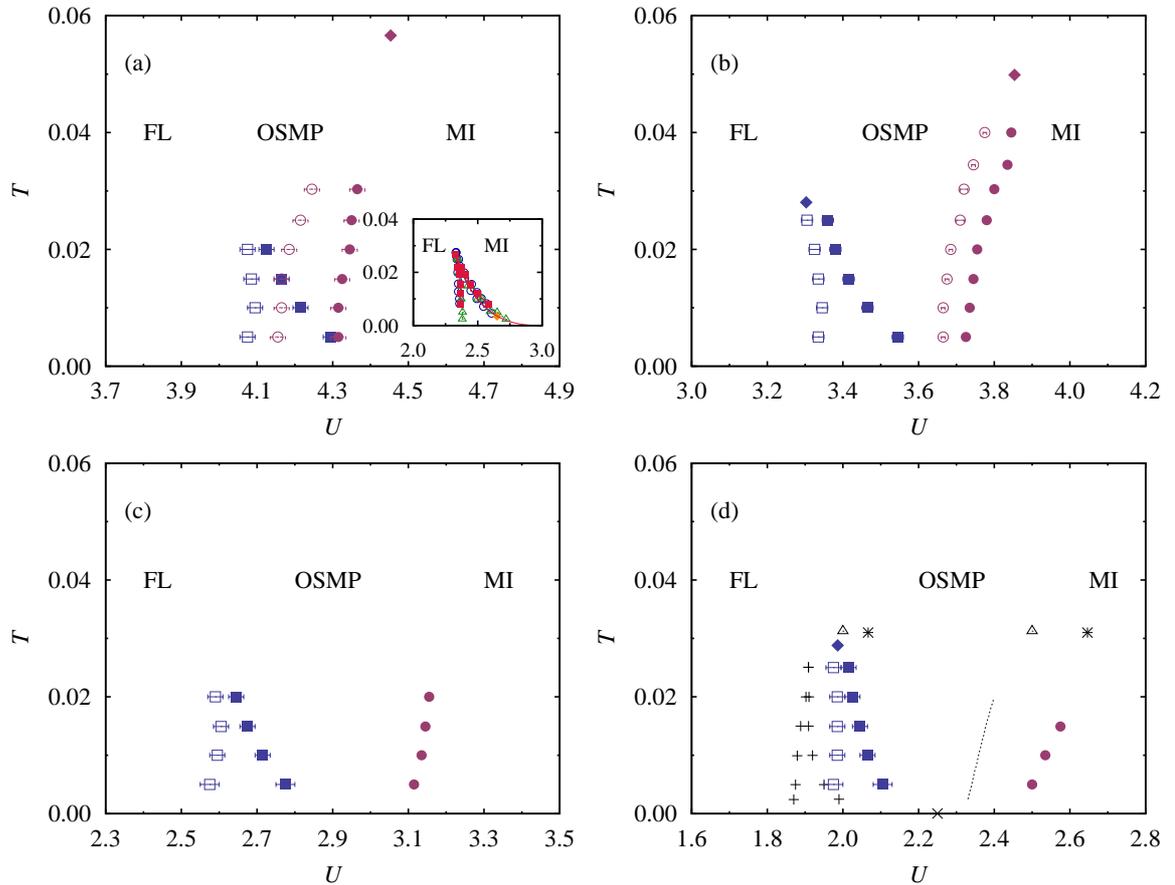}
	\caption{(Color online) Phase diagrams on the plane of temperature $T$
		and interaction strength $U$ for (a) $J=U/32$, (b) $J=U/16$, (c) $J=U/8$, and (d) $J=U/4$.
		In (a) and (b), open and filled symbols correspond to lower and upper
		transition interaction strengths, respectively.  The transitions
		for narrow and wide orbitals are denoted by (blue) squares and (red) circles, respectively.
		The inset in (a) exhibits transition interaction strengths of the single-orbital Hubbard model.
		In (c) and (d) (blue) open squares and (blue)
		filled squares represent lower and upper transition interaction strengths, respectively, of the narrow orbital transition and (red) filled circles the phase boundary within the crossover region between the OSMP and the MI phase.
		The (black) pluses [+] as well as the dashed line represent the result of finite-temperature ED~(Ref.~\onlinecite{Liebsch2005}) while that of zero-temperature ED~(Ref.~\onlinecite{Koga2004}) is represented by (black) crosses [$\times$].
		The results from the HF-QMC method in Refs.~\onlinecite{Knecht2005} and \onlinecite{Liebsch2004} are marked by a (black) open triangle and {asterisks} [$\ast$], respectively.
The diamonds represent the critical end points of the corresponding first-order transitions. }
\label{fig:TUphase}
\end{figure*}

We employ the DMFT combined with the CTQMC method through the hybridization expansion algorithm~\cite{Hafermann2012,Haule2007,Gull2011}.
Typically, the statistical sampling of $10^8$ Monte Carlo steps is performed,
which turns out to be sufficient for statistically reliable numerical results.

\section{RESULTS}
\label{sec:results}

\subsection{Finite-Temperature Phase Diagram}

The main result of this work is
the emergence of a slope-reversed Mott transition
accompanied by the drastic change in the behavior of
finite-temperature phase transitions, which is driven by the variation in Hund's coupling.
Figures~\ref{fig:TUphase} shows phase diagrams on the temperature versus
interaction strength plane for various Hund's coupling strengths.
In the presence of orbital degrees of freedom, generally, we have two successive phase transitions, one from the Fermi-liquid (FL) phase to the OSMP and the other from the OSMP to the MI phase.
The transition between the FL phase and the OSMP inherits the shape and energy scale of the coexistence region in the single-orbital model.
In Fig.~\ref{fig:TUphase}(a) and (b), on the other hand, the coexistence region of
the OSMP-to-MI phase transition is quite interesting.
First of all, the slope of the phase-transition line is opposite to that in the single-orbital case shown in the inset of Fig.~\ref{fig:TUphase}(a).
The slope-reversed Mott transition was reported in the two-dimensional systems and its origin was attributed to spatial modulations~\cite{Park2008}.
Here it is noted that our system is an infinite-dimensional one without any spatial
fluctuations.
We can also find that the critical temperature
associated with the {slope-reversed} transition is considerably enhanced.

The effects of Hund's coupling are rather drastic on the slope-reversed Mott transition.
When we increase the Hund's coupling strength, 
the slope-reversed Mott transition becomes a finite-temperature crossover,
implying a continuous transition at zero temperature.
A similar change in the zero-temperature transition was
reported in an effective low-energy model~\cite{Costi2007};
our result reveals that the zero-temperature result reflects the change from
the slope-reversed transition to a crossover at finite temperatures.
In addition, the region of the OSMP, which is present between the two
transitions, becomes wider for larger Hund's coupling strength,
from which we can infer that Hund's coupling plays the role of a `band
decoupler'~\cite{deMedici2011}.
It is also found that for very small Hund's coupling strength, $J=U/32$, the coexistence regions of the two transitions overlap significantly with each other.
In Fig.~\ref{fig:TUphase}(d) we also plot the existing results obtained from
exact diagonalization (ED)~\cite{Liebsch2005,Koga2004} and Hirsch-Fye quantum Monte Carlo (HF-QMC)~\cite{Knecht2005,Liebsch2004},
which are reasonably consistent with our numerical results.

The reversed slope of the phase-transition line is a distinctive feature.
In a conventional Mott transition the localized MI phase dominates the
itinerant FL phase in the region of high temperatures near the phase transition;
this is mainly due to the extensive entropic contribution of the MI phase
compared with the very small ground-state entropy in the FL phase.
Similarly to the slope-reversed transition in two-dimensional systems,
the origin of which is the significant entropy reduction of the MI phase by the short-range
correlations~\cite{Park2008},
the slope-reversed Mott transition in the two-orbital system
can be understood in terms of the entropy of the MI phase: It is expected to reduce
considerably through ferromagnetic correlations between electrons in different orbitals by Hund's coupling.
Another important aspect in the two-orbital system is that instead of the FL phase,
the OSMP competes with the MI phase near the transition.
The OSMP, in which electrons are partly localized, has higher entropy
than the FL, and accordingly it is more likely to dominate the MI
phase at high temperatures to yield the reversed slope of the transition line.
We will give a detailed analysis in a later subsection, where the temperature-dependence of the energy density is discussed.

\subsection{Spectral Function and Self-Energy}

\begin{figure}
	\centering
	\includegraphics[width=0.45\textwidth]{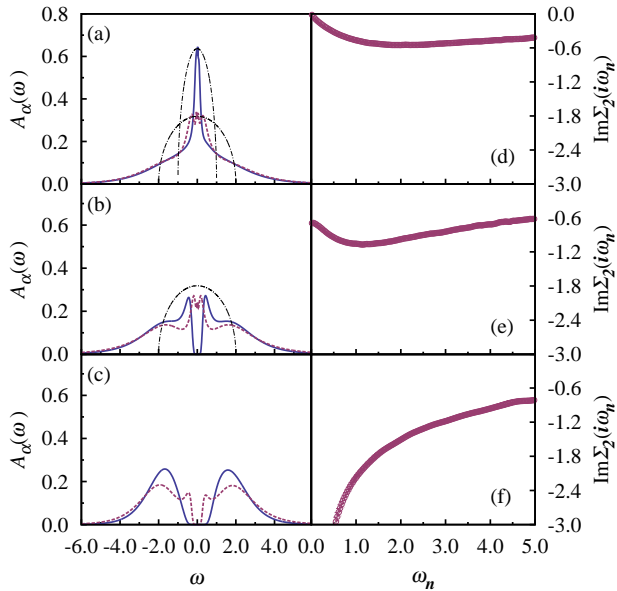}
	\caption{(Color online) Spectral functions calculated via the maximum entropy
	method [(a) to (c)] and imaginary-part self-energies of wide-orbital electrons
[(d) to (f)] at $T=1/200$ for $J/U=1/16$. From top to bottom, the interaction strength
corresponds to $U= 3.00$ [(a) and (d)], $3.60$ [(b) and (e)], and $4.00$ [(c) and (f)].
In (a) to (c), the (blue) solid and the (red) dashed lines represent the spectral
functions of narrow and wide orbitals, respectively. For comparison, the noninteracting density of states, marked with the (black) dot-dashed line, is also shown in (a) both orbitals and (b) wide orbital.}
	\label{fig:A}
\end{figure}

The local spectral function of each orbital,
which can be evaluated via an analytic continuation to the real-frequency domain
by the maximum entropy method (MEM),
characterizes conveniently the feature of each phase in the phase diagram.
In the left panel of Fig.~\ref{fig:A}, the spectral functions of three different phases are shown
for $J=U/16$.
In the FL phase with $U=3.0$, the spectral function exhibits clearly a coherent
peak, which satisfies the Luttinger theorem.
On the other hand, the coherent peak disappears and Mott gaps develop for both
orbitals in the MI phase.
For the intermediate interaction strength corresponding to OSMP, the narrow
orbital is gapped while the wide one still remains itinerant.
It is remarkable that the spectral function of the wide orbital deviates substantially
from the noninteracting DOS at the Fermi level.
The violation of the Luttinger theorem implies the finite lifetime of wide-orbital electrons at the Fermi level.
The finite-scattering amplitude of the wide-orbital electron at the Fermi level can
be verified by the finite offset in the imaginary-part of the self-energy, as shown in
Fig.~\ref{fig:A}(e).
Similar evidences
were also reported for the non-Fermi-liquid nature of the OSMP
which crosses over to the MI phase~\cite{Biermann2005,Liebsch2006}.

\subsection{Local Magnetic Moments}
\begin{figure}
	\centering
	\includegraphics[width=0.45\textwidth]{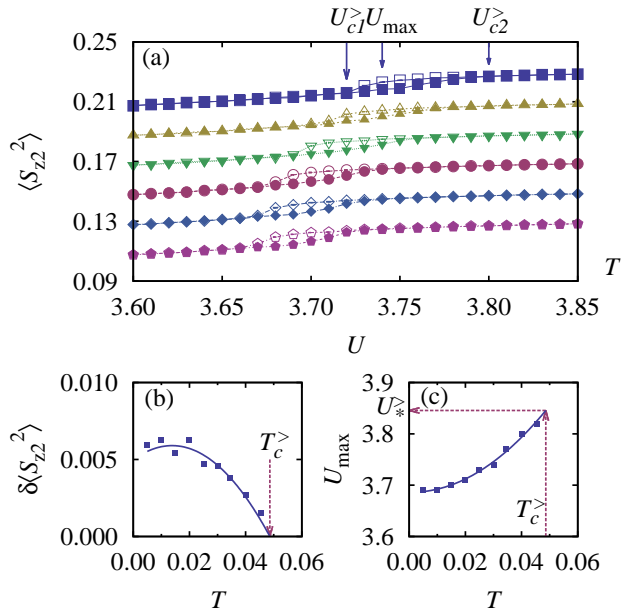}
	\caption{(Color online) (a) Squared magnetic moments of wide-orbital
	electron as a function of the interaction strength at various temperatures
for $J/U=1/16$. From top to bottom, corresponding temperatures are
$T=1/33, 1/40, 1/50, 1/67, 1/100$, and $1/200$.
For better comparison,
the data for $T=1/40, 1/50, 1/67, 1/100$, and $1/200$ are shifted downward by $0.02, 0.04, 0.06, 0.08$, and $0.10$, respectively. Lines are guides to the eye.
(b) Maximum differences of the squared magnetic moments between two solutions in the coexistence region at given temperatures. The solid line corresponds to the least-square fit of the data. 
The critical temperature $T_c^>$ is estimated by the $T$-axis cut of the
extrapolated line, which is denoted by the (red) dashed arrow.
(c) Interaction strength $U_{\rm{max}}$ at which the difference reaches the maximum.
The critical interaction strength $U_{*}^>$ is estimated by the the extrapolation
of the least-square fit [(blue) solid line] to the critical temperature [(red) vertical dashed arrow].}
	\label{fig:SzSq}
\end{figure}

The first-order transition between the OSMP and the MI phase is demonstrated by the
hysteresis behavior of physical quantities such as the local magnetic moment.
In Fig.~\ref{fig:SzSq}(a) we plot the local magnetic moment of the wide orbital as a
function of $U$ for different temperatures.
As the interaction strength is increased, electrons become more localized and the average local moment increases monotonically.
Over a finite region of the interaction strength we can observe the hysteresis of the local
spin magnetic moment, which implies the coexistence of the two phases.
As shown in Fig.~\ref{fig:SzSq}, we can estimate two transition interaction
strengths $U_{c1}^>$ and $U_{c2}^>$ from the minimum and the maximum values of $U$, respectively,
showing the coexistence.
The coexistence region shifts to the stronger interaction region with the increase of
the temperature, resulting in the reversed slope of the phase-transition line.

Using the above hysteresis, we can also estimate the position of the critical end point of the slope-reversed Mott transition.
From the numerical data, we obtain
the maximum difference of the local moments for the two solutions (MI phase and OSMP)
in the coexistence region
\begin{equation}
\delta\langle S^2_{z2}\rangle
\equiv
\mathop{\rm Max}_U \left[\langle S^2_{z2}\rangle_{\rm{MI}} - \langle
S^2_{z2}\rangle_{\rm{OSMP}}\right]
\end{equation}
at each temperature.
In the plot of $\delta\langle S^2_{z2}\rangle$
as a function of $T$, the $T$-axis cut gives the critical temperature $T_c^>$, as
shown in Fig.~\ref{fig:SzSq}(b).
The hysteresis data provide the interaction strength $U_{\rm {max}}$, where
$\delta\langle S^2_{z2}\rangle$ reaches the maximum, and
the extrapolated value of $U_{\rm max}$ to $T=T_c^>$ gives the interaction
strength $U_*^>$ of the critical end point. [See Fig.~\ref{fig:SzSq}(c).]
We have thus determined the location of the critical end points for both
first-order transitions, which are plotted in Fig.~\ref{fig:TUphase}.

\subsection{Origin of Slope-Reversed Mott Transitions}
\begin{figure}
	\centering
	\includegraphics[width=0.45\textwidth]{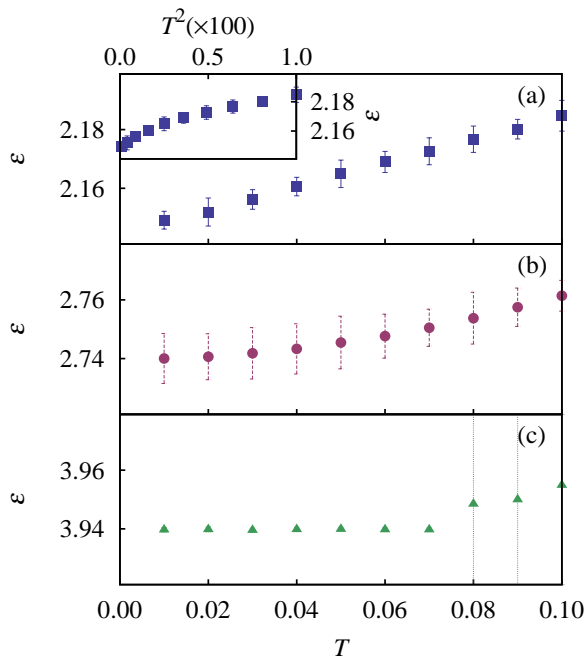}
	\caption{(Color online) Total energy density as a function of temperature 
with $J/U=1/16$
for three different interaction strengths 
		$U=$ (a) $3.0$, (b) $3.6$, and (c) $5.0$,
which correspond to the FL phase, the OSMP, and the MI phase, respectively.
The inset in (a) exhibits $T^2$ behavior of the Fermi-liquid phase at low temperatures.
	}
	\label{fig:E}
\end{figure}

The investigation of the temperature-dependence of the total energy density 
sheds light on the origin of the slope-reversed transitions between the OSMP and MI phase.
Figure~\ref{fig:E} represents the total energy density $\varepsilon$
as a function of temperature for the three phases,
where $\varepsilon$ is defined to be
\begin{equation}
	\varepsilon \equiv \frac{1}{N} \left\langle \mathcal{H} +
\mu\sum^{}_{i\alpha\sigma}\hat{n}_{i\alpha\sigma} \right\rangle~
\end{equation}
with $N$ being the number of lattice sites.

In the localized MI phase the total energy density is nearly constant, which reflects the fact that the entropy is insensitive to the temperature at low temperatures.
On the other hand, the itinerant FL phase gives a monotonic increase in the total energy density with the increase of the temperature.
As shown in the inset, the increasing behavior is consistent with $T^2$ behavior
at low temperatures.
Interestingly, in the OSMP the total energy density also increases as the temperature is increased as in the FL phase.
Such an increase in $\varepsilon_{\rm OSMP}$ makes the OSMP more favorable
compared with the MI phase through the additional contribution to the entropy at finite temperatures. 
Here it is noted that
the temperature dependence in the OSMP shows the superlinear behavior,
$\varepsilon_{\rm OSMP}(T) \approx A T^\gamma$ with $\gamma>1$.

We can also see that the residual entropy of the ground state is $\ln2$ per
site in the OSMP and the MI phase. (Note that the Boltzmann
constant has been absorbed in the temperature $T$.)
In the OSMP, only the electrons in narrow orbitals are localized and the degree
of freedom 
for their spins gives the residual entropy $\ln 2$.
In the MI phase,
on the other hand, electrons in both narrow and wide orbitals are localized.
Nevertheless,
the Ising-type Hund's coupling makes the ground state of the local Hamiltonian
be still two-fold degenerate, composed of $|\uparrow;\uparrow\rangle$ and
$|\downarrow;\downarrow\rangle$, where $|\alpha;\beta\rangle$ describes the
state with a spin-$\alpha$ electron in the narrow orbital and a spin-$\beta$
electron in the wide orbital.

\begin{figure}
	\centering
	\includegraphics[width=0.5\textwidth]{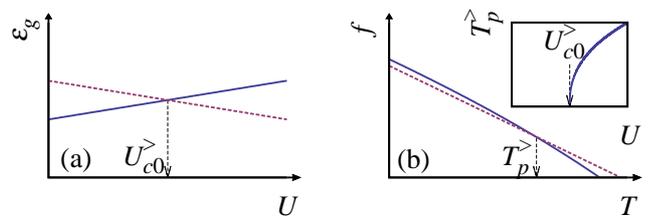}
	\caption{(Color online) 
		(a) Schematic representation of the ground-state energy
		densities of the OSMP and the MI phase, which are represented by the (blue) solid and the (red) dashed lines, respectively.
		$U_{c0}^>$ denotes the transition interaction strength at zero temperature.
		(b) Variations of the free-energy densities of the OSMP [(blue)
		solid line] and the MI phase [(red) dashed line] for $U>U_{c0}^>$ as the temperature increases. 
		Phase transition from the MI phase to the OSMP occurs at $T_p^>$
		only for $U>U_{c0}^>$.
		The inset displays the resulting transition temperature as a function of $U$, which reproduces well the slope-reversed Mott phase transition.
	}
	\label{fig:F}
\end{figure}

Based on these results, we can construct a generic phase boundary between the OSMP and the MI phase.
Suppose 
a zero-temperature quantum phase transition between the OSMP and the MI phase takes place at 
$U=U_{c0}^{>}$, where the ground state energies of two phases cross, as illustrated
in Fig.~\ref{fig:F}(a). 
We use the relation 
\begin{equation}
	s(T) = s(T=0) + \int_{0}^{T}\frac{dT'}{T'}~\frac{d\varepsilon(T')}{dT'}~
	\label{eqn:entropy}
\end{equation}
to estimate the entropy density $s(T)$ at low temperatures.
The resulting temperature dependencies of the free-energy density for the two
phases are given in the form
\begin{eqnarray} \label{eq:f_OSMP}
	f_{\mathrm{OSMP}} &=& \varepsilon_{g,\mathrm{OSMP}} - T\ln 2 - \frac{A \gamma}{\gamma-1}T^\gamma~,\\
	f_{\mathrm{MI}} &=& \varepsilon_{g,\mathrm{MI}} - T\ln 2~.
\end{eqnarray}
Both phases have the same residual entropies while the OSMP has additional
free-energy gain,
shown in
the third term of Eq.~(\ref{eq:f_OSMP}).
This contribution originates from the superlinear temperature dependence of the energy density and the corresponding entropy gain in the OSMP at finite temperatures. 
For $U<U_{c0}^>$, $f_{\rm OSMP}$ is always lower than $f_{\rm MI}$
since $\varepsilon_{g,\mathrm{OSMP}}<\varepsilon_{g,\mathrm{MI}}$. 
For $U>U_{c0}^>$, on the other hand,
$\varepsilon_{g,\mathrm{OSMP}}>\varepsilon_{g,\mathrm{MI}}$ and there occurs a phase
transition at the temperature $T_p^>$ given by
\begin{equation}
	T_p^> = \left[\frac{\gamma-1}{A \gamma}(\varepsilon_{g,\mathrm{OSMP}}-\varepsilon_{g,\mathrm{MI}})\right]^{1/\gamma}.
\end{equation}
Below $T_p^>$ the MI phase has lower free energy while the increase of temperature
above $T_p^>$ induces a transition to the OSMP phase.
Near the zero-temperature transition interaction strength $U_{c0}^>$, the ground-state energy
difference $\varepsilon_{g,\mathrm{MI}}-\varepsilon_{g,\mathrm{OSMP}}$ is
expected to be linearly proportional to $U - U_{c0}^>$, resulting in the following dependence
of the transition temperature $T_p \sim (U-U_{c0}^>)^{1/\gamma}$.
The inset of Fig.~\ref{fig:F}(b) represents a generic phase transition line between
the OSMP and the MI phase, which turns out to be slope-reversed.
The resulting phase transition line also reproduces well
the sublinear dependence of $T_p^>$ on $U-U_{c0}^>$, which is observed in
Figs.~\ref{fig:TUphase}(a) and (b).

\subsection{Effects of Hund's Coupling}

\begin{figure}
	\centering
	\includegraphics[width=0.45\textwidth]{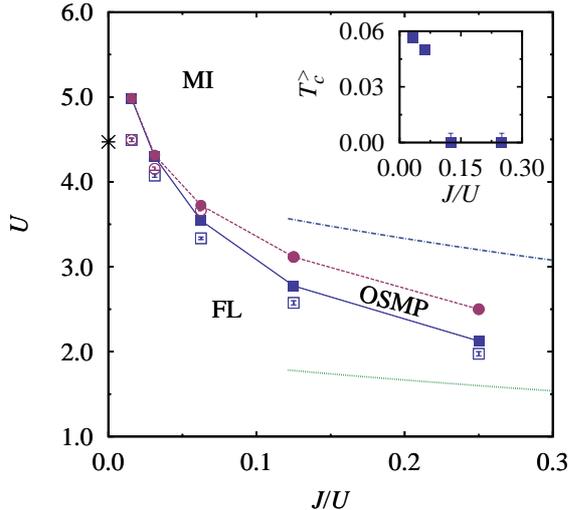}
	\caption{(Color online) Phase diagram on the plane of the interaction
	strength $U$ and the Hund's coupling strength $J/U$ at the temperature $T=1/200$.
	Filled and open (blue) squares indicate upper $(U_{c2}^<)$ and lower
	$(U_{c1}^<)$  transition interaction
strengths, respectively, of the narrow-orbital transition.
For $J/U<0.1$, the (red) filled and open circles represent upper $(U_{c2}^>)$
and lower $(U_{c1}^>)$ 
transition interaction strengths of the wide-orbital transition.
For $J/U>0.1$, the crossover points between the OSMP and the MI phase are marked by the (red) filled circles. Lines are guides to the eye.
The transition interaction strengths for the narrow and the wide  orbitals, which are estimated in the limit $J\gg t_\alpha$, are represented by 
(green) dotted and (blue) dash-dotted lines, respectively.
The transition interaction strength for $J=0$ is also marked by
a (black) asterisk. 
Inset: Critical temperature of the wide-orbital transition versus $J/U$.
}
	\label{fig:UJphase}
\end{figure}

In Fig.~\ref{fig:UJphase}
we summarize the effects of Hund's coupling on the transitions
by plotting various transition interaction strengths versus $J/U$ at
$T=1/200$, which is the lowest temperature considered.
For $J/U=1/64$, the system appears to undergo a single transition without
the OSMP.
For larger values of $J/U$, we can observe two separate transitions, and the region of OSMP
expands gradually with the increase of $J/U$.
It is also notable that
the critical interaction strengths associated with both orbitals tend to decrease as Hund's coupling grows.

	Following the Hubbard criterion for the Mott transition, which is extended for the multi-orbital models~\cite{Koga2004,Georges2013},
	we can simply estimate critical interaction strength.
	In the extremely localized atomic limit $(t_\alpha=0)$, the charge
	excitation gap is given by
	\begin{eqnarray}
		\Delta_{\mathrm{atom}} &=& \left[E_g(N+1) - E_g(N)\right] - \left[E_g(N) - E_g(N-1)\right]\nonumber\\
		&=& (1+J/U)U~,
	\end{eqnarray}
	where $E_g(n)$ is the ground-state energy with $n$ electrons.
	The gap is reduced by the introduction of the kinetic energy, and 
	at the critical interaction strength the reduced gap vanishes:
	\begin{equation}
		0= \Delta_{\mathrm{atom}} - \widetilde{W}~,
	\end{equation}
	where $\widetilde{W}$ is the estimate of the average kinetic energy.
	For $J\ll t_\alpha$ the charge excitations in both orbitals are
	hybridized with each other.
	Accordingly, the single Mott transition arises in this limit.
	The charge excitations in both orbitals make contribution to the kinetic
	energy, yielding the estimate  
	$2\sqrt{D_{1}^2+D_{2}^{2}}$ for the average kinetic energy; this results in the enhanced critical interaction strength.
	In the opposite limit $J\gg t_\alpha$, in contrast, orbital
	fluctuations are strongly suppressed and charge excitations in the two
	orbitals are not hybridized with each other.
	The average kinetic energy of the orbital $\alpha$
	reduces to the bare bandwidth $2D_\alpha$, leading to
	the two transition interaction strengths 
	\begin{eqnarray}
		U_c^< &=& \frac{2D_1}{1+J/U}~,
		\\
		U_c^> &=& \frac{2D_2}{1+J/U}~,
	\end{eqnarray}
	which generally decrease with $J/U$.
	
	The above estimates of the transition interaction strengths are qualitatively consistent with our numerical data.
	The Hund's coupling decouples the excitations in two different orbitals,
	and the transition interaction strengths of the two orbitals begin to be separated as the Hund's coupling strength is raised.
	The corresponding OSMP region becomes enlarged in the phase diagram.
	Hund's coupling thus plays the role of `band decoupler'. 
	The interpolation between $J=0$ and the limit of $J\gg t_\alpha$ clearly
	shows that the transition interaction strength is a decreasing function of $J/U$.
	This is a characteristic of the half-filled system and different
	behaviors in general fillings were reported in several works~\cite{Fresard1997,Lombardo2005,Werner2009,deMedici2011,Georges2013}.

The inset of Fig.~\ref{fig:UJphase} shows that the critical temperature $T_c^>$ of the wide-orbital first-order transition reduces as $J/U$ is increased.
Above a certain value of $J/U$,
which turns out to be between $1/16$ and $1/8$,
we cannot find the transition down to $T=1/200$, the lowest temperature considered,
only to observe crossover phenomena.
We presume that the critical temperature of the Mott transition continues to diminish as $J/U$ increases and eventually becomes zero between $J/U=1/16$ and $1/8$;
this explains the drastic change in
transition nature from the first-order to crossover.

\subsection{Spin-Spin Correlation Function}
\begin{figure}
	\centering
	\includegraphics[width=0.49\textwidth]{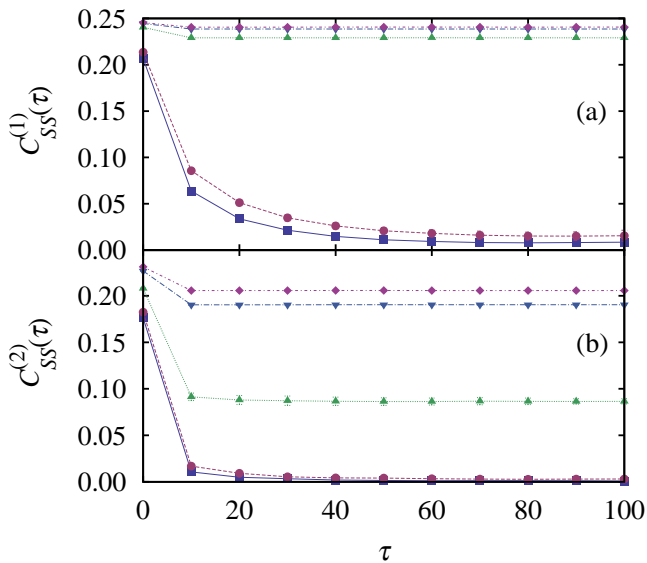}
	\caption{(Color online) Imaginary-time dependence of the spin-spin
		correlation function at $T=1/200$ with $J=U/16$ for (a) the
		narrow and (b) the wide orbitals.
	From bottom to top, the interaction strength is given by $U=3.1$, $3.3$, $3.6$, $3.8$, and $4.0$~.}
	\label{fig:SzSz}
\end{figure}

We next investigate the spin-spin correlation function
\begin{equation}
	C_{SS}^{(\alpha)}(\tau) \equiv \langle S^z_\alpha(\tau)S^z_\alpha(0)\rangle~
\end{equation}
for orbital $\alpha=1,2$.
The spin-spin correlation function can give a signal for the formation of the frozen local magnetic moment.
The long-term memory in the correlation function is proportional to the magnitude of frozen moments.
Figure~\ref{fig:SzSz} represents the spin-spin correlation function of the
narrow and the wide orbitals at $T=1/200$ for $J=U/16$ and various interaction strengths.

In the FL phase, $C_{SS}^{(\alpha)}(\tau)$ for both orbitals shows $1/\tau^{2}$
scaling for imaginary time $\tau$ sufficiently far from both $0$ and
$\beta$. 
In the OSMP, however, we can find the formation of
the frozen local moment in the itinerant wide orbital $(\alpha=2)$, which
exhibits the long-term memory in $C_{SS}^{(2)}(\tau)$. (See the data for $U=3.6$.)
In comparison with the moment of the narrow orbital, that of the wide orbital is not fully developed in magnitude.
Via the second transition, the frozen moment of wide orbital is fully developed
as well, and the system enters the MI phase.
In the OSMP, we presume that not only the local moment of the narrow orbital but
also the frozen moment of the wide orbital can enhance the scattering amplitude
of itinerant electrons in the wide orbital, which is observed in Fig.~\ref{fig:A}(e).
This itinerant phase in the wide orbital is a simple example of `frozen-moment' metal at half filling.
Similar phases were observed at other fillings~\cite{Werner2008,Hafermann2012}.

\begin{figure}
	\centering
	\includegraphics[width=0.49\textwidth]{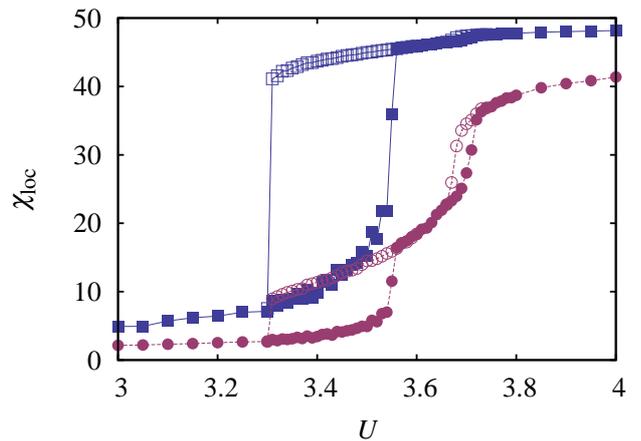}
	\caption{(Color online) Local spin susceptibility as a function of
		interaction strength at $T=1/200$ with $J=U/16$.
		The data for the narrow and the wide orbitals are marked by (blue) squares and (red) circles, respectively.
	}
	\label{fig:chi}
\end{figure}
The local spin susceptibility, defined to be 
\begin{equation}
	\chi^{(\alpha)}_{\mathrm{loc}} = \int_{0}^{\beta}d\tau~\langle
	S^{z}_{\alpha}(\tau)S^{z}_{\alpha}(0)\rangle~,
\end{equation}
is also shown in Fig.~\ref{fig:chi}.
Two successive first-order transitions are clearly observed. 
The intermediate OSMP has moderate values of $\chi^{(2)}_{\mathrm{loc}}$,
which provides another signature of frozen local moment.
Such two-stage saturation of the local susceptibility was reported earlier
and Hund's coupling was also emphasized as an origin of the formation of the local moments in itinerant components~\cite{DeMedici2009a}.

\section{CONCLUSION}
\label{sec:conclusion}
We have found the slope-reversed Mott transition in the two-orbital Hubbard model with Ising-type Hund's coupling, in which two orbitals have different bandwidths.
The reversed slope of the phase-transition line between the OSMP and the MI phase
can be understood in terms of entropy contributions which are closely related to the anisotropy in the Hund's coupling.
The analysis of the temperature dependence of the energy densities
together with the residual entropy has given a successful explanation of a
generic slope-reversed transition between the OSMP and the MI phase.
We have also observed drastic changes in transition nature between the OSMP and the MI phase as the Hund's coupling strength is varied.
As the Hund's coupling strength increases, the first-order transition turns into a finite-temperature crossover, implying a quantum phase transition at zero temperature.
Such a drastic change in the transition nature is apparently induced by the
diminishing critical temperature of the first-order transition between the OSMP and the MI phase.
Finally, the frozen local moments have been observed for the wide orbital in the OSMP.

\section*{ACKNOWLEDGEMENT}
This work was supported by the National Research Foundation of Korea through
Grant No. 2013R1A1A2007959 (A.J.K and G.S.J.) and Grant No.
2012R1A2A4A01004419 (M.Y.C.).

\bibliography{myref}
\end{document}